\documentclass[journal]{IEEEtran}
\ifCLASSINFOpdf

\else
\fi
\usepackage{graphicx}

\usepackage{color}
\usepackage{cite}
\usepackage{braket}
\usepackage{lineno}
\begin{document}
%
\title{Uncertainty evaluation of an $^{171}$Yb \\optical lattice clock at NMIJ}
%
%
%

\author{Takumi Kobayashi, Daisuke Akamatsu, Yusuke Hisai, Takehiko Tanabe, Hajime Inaba, Tomonari Suzuyama, Feng-Lei Hong, Kazumoto Hosaka, and Masami Yasuda
\thanks{T. Kobayashi, D. Akamatsu, T. Tanabe, H. Inaba, T. Suzuyama, K. Hosaka, and M. Yasuda are with the National Metrology Institute of Japan (NMIJ), National Institute of Advanced Industrial Science and Technology (AIST), 1-1-1 Umezono, Tsukuba, Ibaraki 305-8563, Japan (email: takumi-kobayashi@aist.go.jp)}
\thanks{Y. Hisai and F.-L. Hong are with the Department of Physics, Graduate School of Engineering Science, Yokohama National University, 79-5 Tokiwadai, Hodogaya-ku, Yokohama 240-8501, Japan}
\thanks{This work was supported by the Japan Society for the Promotion of Science (JSPS) KAKENHI Grant Number 15K05238, 17K14367, 17H01151, and 18H03886.
}
}

%
%

\markboth{Journal of \LaTeX\ Class Files}%
{Shell \MakeLowercase{\textit{et al.}}: Bare Demo of IEEEtran.cls for IEEE Journals}
%



\maketitle

\begin{abstract}
We report an uncertainty evaluation of an $^{171}$Yb optical lattice clock with a total fractional uncertainty of $3.6\times10^{-16}$, which is mainly limited by the lattice-induced light shift and the blackbody radiation shift. Our evaluation of the lattice-induced light shift, the density shift, and the second-order Zeeman shift is based on an interleaved measurement where we measure the frequency shift using the alternating stabilization of a clock laser to the $\mathrm{6s^{2}\,^{1}S_{0}-6s6p\,^{3}P_{0}}$ clock transition with two different experimental parameters. In the present evaluation, the uncertainties of two sensitivity coefficients for the lattice-induced hyperpolarizability shift $d$ incorporated in a widely-used light shift model by RIKEN and the second-order Zeeman shift $a_{\mathrm{Z}}$ are improved compared with the uncertainties of previous coefficients. The hyperpolarizability coefficient $d$ is determined by investigating the trap potential depth and the light shifts at the lattice frequencies near the two-photon transitions $\mathrm{6s6p^{3}P_{0}-6s8p^{3}P_{0}}$, $\mathrm{6s8p^{3}P_{2}}$, and $\mathrm{6s5f^{3}F_{2}}$. The obtained values are $d=-1.1(4)$ $\mathrm{\mu}$Hz and $a_{\mathrm{Z}}=-6.6(3)$ Hz/mT$^{2}$. These improved coefficients should reduce the total systematic uncertainties of Yb lattice clocks at other institutes. 
\end{abstract}


%
\IEEEpeerreviewmaketitle

\section{Introduction}
%
%
%
%
\IEEEPARstart{O}{ptical} clocks based on narrow optical transitions have demonstrated fractional uncertainties at the $10^{-18}$ level \cite{Chou2010,Bloom2014,Ushijima2015,Nicholson2015,Huntemann2016,Nemitz2016}, surpassing the Cs fountain microwave clocks that realize the second in the International System of Units (SI). This fact has stimulated discussions regarding a redefinition of the SI second using optical clocks \cite{Gill2011,Riehle2018,Hong2016}. Accurate optical clocks are expected to contribute to tests of fundamental physics \cite{Godun2014,Huntemann2014,Derevianko2014}, relativistic geodesy \cite{Yamaguchi2011,Takano2016, Koller2017,Yasuda2017,Grotti2018}, and the generation of a stable time scale \cite{Grebing2016,Lodewyck2016,Hachisu2018}. Optical lattice clocks are one of promising optical clocks for a future definition of the second. A large number ($\sim10^{3}-10^{4}$) of neutral atoms are trapped in an optical lattice operated at a ``magic" wavelength where the lattice-induced ac Stark shift (light shift) is cancelled \cite{Katori2003}. 

Among several atomic species used to realize optical lattice clocks, Yb has been actively studied by several groups around the world. The $\mathrm{6s^{2}\,^{1}S_{0}-6s6p\,^{3}P_{0}}$ clock transition of $^{171}$Yb has been adopted as one of secondary representations of the second by the Comit\'e International des Poids et Mesures (CIPM) \cite{Riehle2018}. A frequency stability at the 10$^{-18}$ level has been achieved at the National Institute of Standards and Technology (NIST) \cite{Schippo2017,Hinkley2013} and RIKEN \cite{Nemitz2016}. Absolute frequency measurement of the clock transition relative to the SI second has been carried out at the National Metrology Institute of Japan (NMIJ) \cite{Kohno2009, Yasuda2012}, NIST \cite{Lemke2009}, the Korea Research Institute of Standards and Science (KRISS) \cite{Park2013,Kim2017}, and Istituto Nazionale di Ricerca Metrologica (INRiM) \cite{Pizzocaro2017}. The frequency ratio of the $^{171}$Yb and $^{87}$Sr optical lattice clocks has been measured at NMIJ \cite{Akamatsu2014}, RIKEN \cite{Takamoto2015,Nemitz2016}, Physikalisch-Technische Bundesanstalt (PTB)/INRiM \cite{Grotti2018}, and the National Institute of Information and Communications Technology (NICT)/KRISS \cite{Fujieda2018}. As regards the frequency ratio measurement, we have recently demonstrated the dual-mode operation of the Sr-Yb optical lattice clock \cite{Akamatsu2018}. 

Although Yb optical lattice clocks have performed extremely well as a candidate for the redefinition of the second, the number of uncertainty evaluation data is small compared with that of Sr lattice clocks. With respect to certain important systematic frequency shifts including the lattice-induced hyperpolarizability shift, the second-order Zeeman shift, the blackbody radiation shift, and the probe-induced light shift, most of the relevant groups have employed the sensitivity coefficients measured only at NIST \cite{Lemke2009,Barber2008,Sherman2012,Beloy2014}. The second-order Zeeman shift coefficient was recently measured at the East China Normal University (ECNU) \cite{Gao2018}, but this measurement does not improve the uncertainty of the coefficient reported by NIST. The accumulation of more data of the uncertainty evaluation is essential for confirming the consistency between different independently developed clocks. 

In this paper, we report the uncertainty evaluation of an $^{171}$Yb optical lattice clock developed at NMIJ. Systematic frequency shifts have been evaluated with a total fractional uncertainty of $3.6\times10^{-16}$. In this evaluation, the uncertainties of two coefficients for the hyperpolarizability shift incorporated in a widely-used light shift model by RIKEN \cite{Katori2015,Nemitz2016} and the second-order Zeeman shift are improved compared with the uncertainties of coefficients reported by RIKEN \cite{Nemitz2016} and NIST \cite{Lemke2009}. Our two improved coefficients should reduce the total systematic uncertainties of Yb clocks at RIKEN \cite{Nemitz2016}, KRISS \cite{Kim2017}, and INRiM \cite{Pizzocaro2017}.

\section{Experimental setup}
\label{experimentalsetup}
\begin{figure}[h]
\begin{center}
\includegraphics[width=8cm,bb=0 50 910 816]{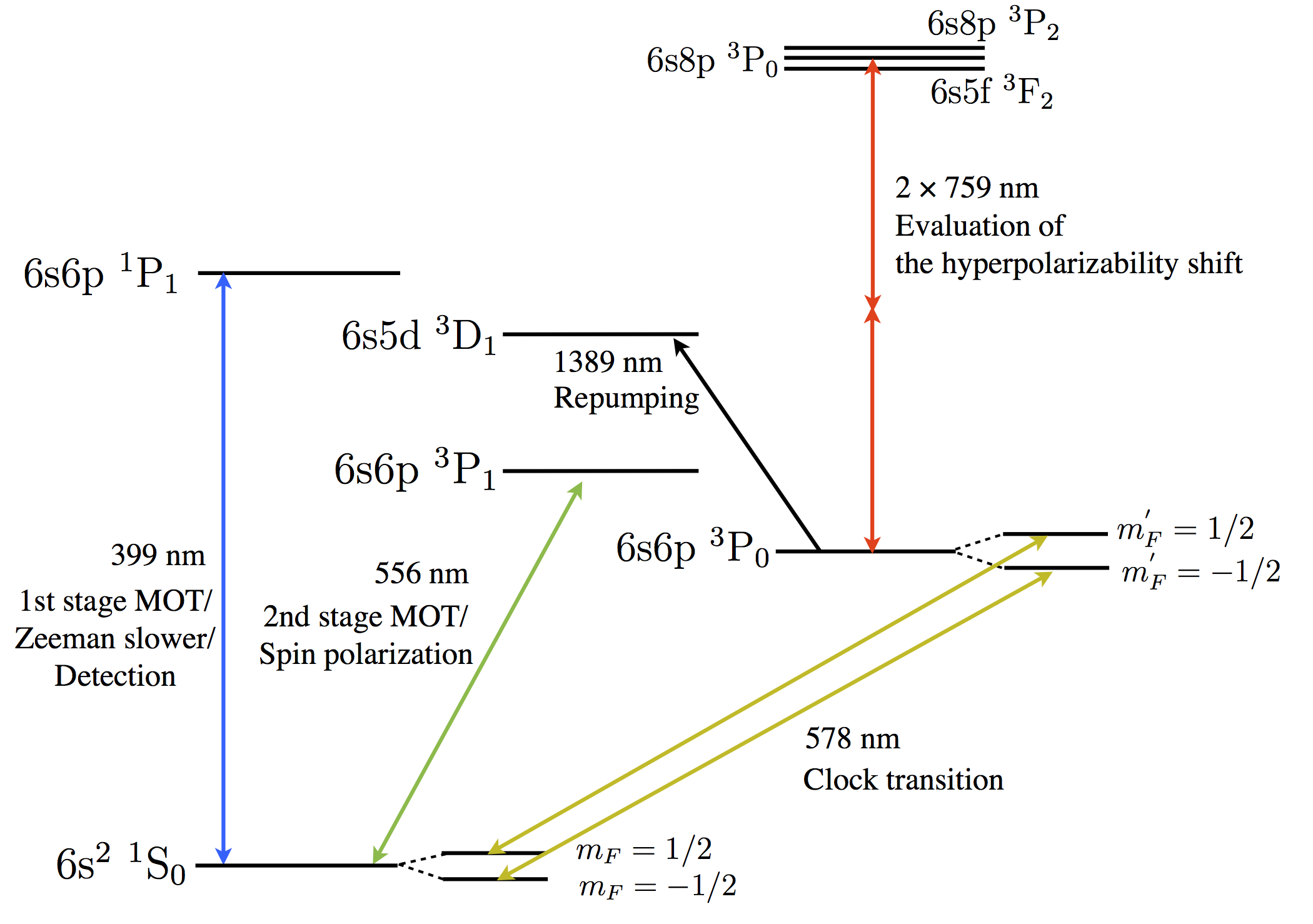}
\end{center}
\caption{Energy diagram of $^{171}$Yb with nuclear spin $I=1/2$ relevant for the optical lattice clock. The Zeeman splittings ($m_{F}=\pm 1/2$) of the states $^{1}$S$_{0}$ and $^{3}$P$_{0}$ are shown. To evaluate the hyperpolarizability shift, the frequency of the optical lattice was tuned to the two-photon resonances $\mathrm{6s6p^{3}P_{0}-6s8p^{3}P_{0}}$, $\mathrm{6s8p^{3}P_{2}}$, and $\mathrm{6s5f^{3}F_{2}}$ (see Sect. \ref{latticelightshiftsection}).}
\label{energydiagram}
\end{figure}

We here provide a detailed description of the $^{171}$Yb optical lattice clock whose relevant energy diagram is shown in Fig. \ref{energydiagram}. The strong dipole-allowed $\mathrm{6s^{2}\,^{1}S_{0}-6s6p\,^{1}P_{1}}$ transition at 399 nm was used for the first-stage magneto-optical trap (MOT), the Zeeman slower, and the detection of trapped atoms. The $\mathrm{6s^{2}\,^{1}S_{0}-6s6p\,^{3}P_{1}}$ intercombination transition at 556 nm was utilized for the second-stage MOT. The atoms were then loaded to an optical lattice operated at the magic wavelength of 759 nm. The atoms were spin polarized using the $\mathrm{^{1}S_{0}-}\mathrm{^{3}P_{1}}$ transition. The $\mathrm{^{1}S_{0}}\mathrm{-^{3}P_{0}}$ clock transition was probed by a clock laser at 578 nm. The excitation probability of the clock transition was deduced using a repumping light at 1389 nm resonant on the $\mathrm{6s6p\,^{3}P_{0}-6s5d\,^{3}D_{1}}$ transition. 

\subsection{Laser systems}
\label{lasersystemsection}
\begin{figure*}[t]
\begin{center}
\includegraphics[width=11cm,bb=0 50 910 816]{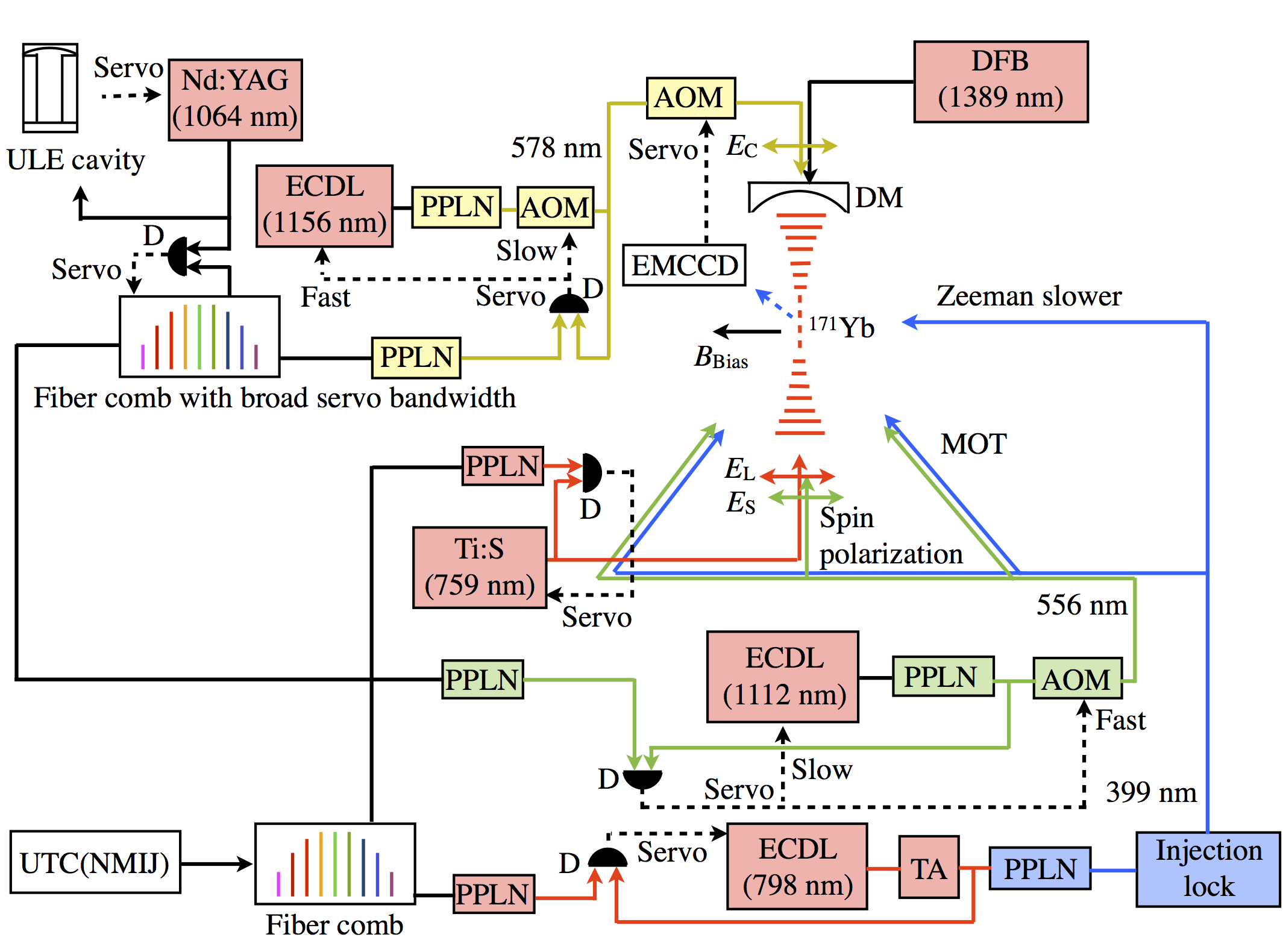}
\end{center}
\caption{Schematic diagram of the experimental setup. ULE: Ultra-low expansion, ECDL: External cavity diode laser, PPLN: Periodically-poled LiNbO$_{3}$, AOM: Acousto-optic modulator, DM: Dichroic mirror, DFB: Distributed-feedback laser, EMCCD: Electron multiplying charged coupled device, Ti:S: Titanium sapphire laser, MOT: Magneto-optical trap beam, TA: Tapered amplifier, UTC(NMIJ): Coordinated universal time of the National Metrology Institute of Japan, D: Detector for beat detection. $E_{\mathrm{L}}$, $E_{\mathrm{C}}$, and $E_{\mathrm{S}}$ denote the electric field vectors for the lattice, clock, and spin-polarization lasers, respectively, which are aligned with the direction of the external bias magnetic field $B_{\mathrm{Bias}}$.}
\label{setupfigure}
\end{figure*}

Figure \ref{setupfigure} shows a schematic diagram of our experimental setup. A 399-nm light for the first-stage MOT and the Zeeman slower was generated using a single-pass periodically poled LiNbO$_{3}$ (PPLN) waveguide for second harmonic generation (SHG) \cite{Kobayashi2016}. A home-made external cavity diode laser (ECDL) emitting at 798 nm with a tapered amplifier (TA) was used as a fundamental input laser. An SHG power of 5 \nolinebreak mW at 399 nm was obtained from the PPLN waveguide. This SHG power was amplified to 200 mW by injection locking of a 399-nm diode laser (Nichia NDV4B16) \cite{Hosoya2015}. The amplified beam was split into two parts for the MOT and Zeeman slower. These two beams were passed through acousto-optic modulators (AOM) to tune the frequencies and sent to a vacuum chamber via single mode optical fibers (not shown in Fig. \ref{setupfigure}). The frequency of the 798-nm ECDL was stabilized to a home-made optical frequency comb based on a mode-locked erbium-doped fiber laser (fiber comb) \cite{Inaba2006} operated at a repetition rate of 122 MHz, which was referenced to the Coordinated Universal Time of the National Metrology Institute of Japan (UTC(NMIJ)). The stabilization was carried out using an electrical delay line \cite{Schunemann1999,Komori2003,Hisai2018} by controlling a piezoelectric transducer (PZT) in the ECDL. 

A 556-nm light for the second-stage MOT was generated by the SHG of a 1112 nm light from a commercially available cateye ECDL (MOGLabs, CEL002). The output of the ECDL was amplified by a semiconductor optical amplifier to a power of 200 mW and coupled to a PPLN waveguide for the SHG. Since the natural linewidth of the  $\mathrm{^{1}S_{0}-^{3}P_{1}}$ transition is 182 \nolinebreak kHz, the linewidth of the laser must be narrow. Therefore, the linewidth of a stable Nd:YAG laser at 1064 nm, which was stabilized to an ultra-low expansion (ULE) cavity via the Pound-Drever-Hall method, was transferred to the 556 \nolinebreak nm laser using a fiber comb with a broad servo bandwidth \cite{Nakajima2010,Iwakuni2012}. The 556-nm laser was phased-locked to the comb by the fast control of the laser frequency with an AOM. A slow frequency drift was compensated for with the PZT of the ECDL. The 556-nm beam was frequency-shifted by another AOM and sent to the vacuum chamber via a single mode fiber (not shown in Fig. \ref{setupfigure}).

The lattice light at 759 nm was provided by a commercially available titanium sapphire (Ti:S) laser (M squared, SolsTiS), which was stabilized using an electrical delay line \cite{Schunemann1999,Komori2003,Hisai2018} to the same fiber comb referenced to UTC(NMIJ). The uncertainty in the determination of the laser frequency was $<100$ \nolinebreak kHz. The laser frequency was monitored with a wavemeter and another fiber comb operated at a different repetition rate (not shown in Fig. \ref{setupfigure}) to detect laser mode hopping. The linearly polarized light passed through a single mode fiber and was then focused on the atoms with a power of  $\sim$0.5 W. The 1/$e^{2}$ waist radius of the lattice beam was estimated to be about $\sim20$ $\mathrm{\mu}$m. The power incident on the atoms was stabilized by detecting a small portion of the lattice power and adjusting the diffraction efficiency of an AOM through which the beam was passed before being coupled to the fiber (not shown in Fig. \ref{setupfigure}).

A 578-nm light for the clock transition was generated by the SHG of an ECDL operated at 1156 nm using a PPLN waveguide \cite{Kobayashi2016JOSAB}. The linewidth transfer scheme described above was also employed for the clock laser to achieve a narrow linewidth \cite{Inaba2013}. The fast frequency control of the ECDL was realized by feeding back an error signal to the injection current of the ECDL. An AOM located after the PPLN was used to provide a slow feedback. The PZT of the ECDL was employed to keep the frequency of the AOM at 80 MHz. The frequency stability of the clock laser was estimated to be about $2\times10^{-15}$ from 1 to 50 s. The clock frequency was scanned by another AOM. The light was sent to the vacuum chamber via a single mode fiber and the power illuminating the atoms was stabilized in the same manner as the lattice light (not shown in Fig. \ref{setupfigure}).  

The repumping 1389-nm light was provided by a commercially produced distributed-feedback (DFB) laser (NTT Electronics, NLK1E5GAA). This light was focused on the atoms with a power of 6 mW, which strongly saturated the $\mathrm{^{3}P_{0}}\mathrm{-^{3}D_{1}}$ transition. This enabled the free-running operation of the DFB laser. 

\subsection{Clock operation}
\label{clockoperationsection}
Yb atoms effused from an oven were decelerated with the Zeeman slower, and trapped and cooled to a millikelvin temperature in the first-stage MOT. The 399-nm beam used for the Zeeman slower had a power of 10 mW and a detuning frequency of $-540$ MHz from the $\mathrm{^{1}S_{0}}\mathrm{-^{1}P_{1}}$ resonance. The MOT was made by three beams and their retro-reflected beams, and operated with a total 399-nm power of 10 mW, a detuning of $-17$ MHz, an 1/$e^{2}$ beam radius of about 4 \nolinebreak mm, and a magnetic field gradient of 0.4 T/m in the strong-gradient axis. The magnetic field was generated by a pair of anti-Helmholtz coils located inside the vacuum chamber. Approximately $3\times10^{6}$ $^{171}$Yb atoms were loaded in the MOT in 1.5 s.

The atoms were further cooled to $\sim40$ $\mathrm{\mu}$K in the second-stage MOT. The 556-nm beam used for the MOT had a total power of 1.2 mW and an 1/$e^{2}$ radius of about 4 mm. The magnetic field gradient was reduced to 0.08 T/m at this stage. For the first 200 ms, the laser frequency was modulated with a spectral bandwidth of 5.5 MHz and detuned by $-6$ MHz to increase the capture velocity range. Then, single mode frequency cooling was performed for 75 ms by reducing the laser power, ceasing the modulation, and decreasing the detuning to 2 MHz.

After the second-stage MOT, the atoms were loaded into a vertically oriented one-dimensional optical lattice. The trap depth was typically 540$E_{\mathrm{r}}$, where $E_{\mathrm{r}}=\hbar^{2} k^{2}/2 m$ is the recoil energy from a lattice photon with a wavenumber $k$, $\hbar$ the reduced Planck constant, and $m$ the mass of the atom. Approximately $6\times10^{2}$ atoms were trapped in the optical lattice with a lifetime of 4 s. The lattice light was turned off during the second-stage MOT, since we observed that this light caused a reduction in the number of atoms trapped in the optical lattice. 

The atoms in the lattice were spin-polarized in the $^{1}$S$_{0}$ state with a magnetic quantum number $m_{F}$ of either $+1/2$ or $-1/2$ by optical pumping on the $\mathrm{^{1}S_{0}}\mathrm{-^{3}P_{1}}$ transition at 556 nm. An external bias magnetic field of 65 $\mathrm{\mu}$T was applied to split the Zeeman components $m_{\mathrm{F}}$ using Helmholtz coils located outside the vacuum chamber. A 556-nm light with its polarization parallel to the bias field induced either of two $\pi$ transitions $m_{F}=+1/2-m^{'}_{F}=+1/2$ or $m_{F}=-1/2-m_{F}^{'}=-1/2$ with a duration of 20 ms, pumping the atoms to the ground dark $m_{F}$ state. The $\pi$ transitions were selected by changing the frequency of the 556-nm light.

\begin{figure}[h]
\begin{center}
\includegraphics[width=9cm,bb=0 50 910 716]{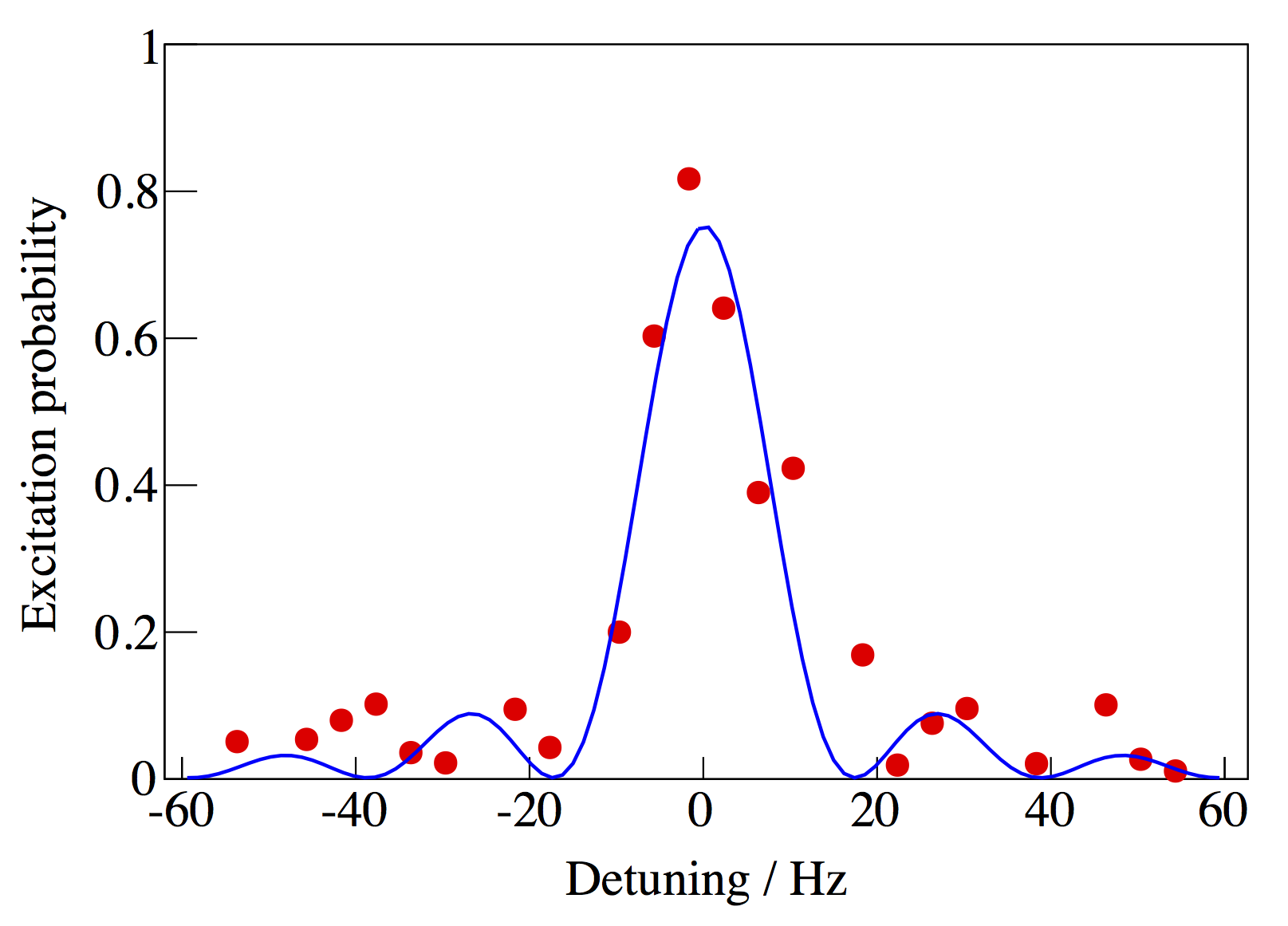}
\end{center}
\caption{Observed spectrum of one Zeeman component of the $\mathrm{^{1}S_{0}}\mathrm{-^{3}P_{0}}$ clock transition with spin-polarized atoms. The solid blue line indicates a Rabi excitation profile calculated for a $\pi$ pulse applied for 50 ms.}
\label{narrowspectrum}
\end{figure}

The $^{1}$S$_{0}$ and $^{3}$P$_{0}$ states relevant to the clock transition were split into two Zeeman components $m_{\mathrm{F}}=\pm1/2$ by applying a magnetic field of 65 $\mathrm{\mu}$T (see Fig. \ref{energydiagram}). The polarizations of the clock and lattice lasers were aligned with the direction of the bias field. A Rabi $\pi$ pulse resonant on the clock transition was applied for typically 50 ms. Two clock $\pi$ transitions were observed with a Zeeman splitting of 275 Hz. Figure \ref{narrowspectrum} shows the observed spectrum of one of the $\pi$ transitions ($m_{F}=-1/2-m^{'}_{F}=-1/2$) with a linewidth of $\sim16$ Hz. The solid blue line indicates the fit of a Rabi excitation profile calculated for a $\pi$ pulse applied for 50 ms. The fit function $P$ is expressed as $P=P_{0}\nu_{\mathrm{R}}^{2}/(\nu_{\mathrm{R}}^{2}+\Delta \nu_{\mathrm{c}}^{2})\sin^{2}(\pi\sqrt{\nu_{\mathrm{R}}^{2}+\Delta\nu_{\mathrm{c}}^2}\Delta t)$, where $P_{0}$ denotes a scaling factor, $\nu_{\mathrm{R}}$ the Rabi frequency, $\Delta \nu_{\mathrm{c}}=\nu_{\mathrm{c}}-\nu_{0}$ the detuning of the clock laser frequency $\nu_{\mathrm{c}}$ from the resonance $\nu_{0}$, and $\Delta t$ the duration of the clock pulse. $P_{0}$ and $\nu_{0}$ were treated as free parameters, while the other parameters were fixed as $\nu_{\mathrm{R}}=10$ Hz and $\Delta t=50$ ms.
 
The excitation probability in Fig. \ref{narrowspectrum} was deduced by counting the numbers of unexcited and excited atoms. To count the number of unexcited atoms in the $^{1}$S$_{0}$ state, a laser-induced fluorescence signal by the $\mathrm{^{1}S_{0}}\mathrm{-^{1}P_{1}}$ transition at 399 nm was first measured using an electron multiplying charge-coupled device (EMCCD). The MOT beam was used to induce the fluorescence. The excited atoms in the $^{3}$P$_{0}$ state were detected by repumping the atoms to the ground $^{1}$S$_{0}$ state, followed by a second fluorescence detection at 399 nm. This repumping was performed by irradiating a repumping light at 1389 nm, which excited the atoms in the $^{3}$P$_{0}$ state to the short-lived $^{3}$D$_{1}$ state. The atoms in this state rapidly decay to the ground state via the $^{3}$P$_{1}$ state. A background signal was generated by stray light and by the fluorescence caused by background atoms. This signal was obtained by a third 399-nm detection without the trapped atoms and subtracted during the deduction of the excitation probability.  

The clock laser probed the two Zeeman components $m_{F}=\pm1/2$ alternately and was stabilized to each component independently. An error signal for each component was obtained by measuring the excitation probabilities of the high- and low-frequency spectral shoulders, and then fed back to the AOM to control the clock frequency to equalize the excitation probabilities. The average frequency of the two components was calculated from the AOM frequencies recorded on a computer, allowing the cancellation of the first-order Zeeman shift and the residual lattice-induced vector light shift \cite{Lemke2009}. The clock cycle time for probing one of the shoulders was 2 \nolinebreak s, and thus a minimum duration of 8 \nolinebreak s was required to obtain the average frequency.

\section{Uncertainty evaluation}
\label{systematicshift}
Table \ref{clockbudget} lists systematic frequency shifts and uncertainties of the Yb optical lattice clock. This section details the uncertainty evaluation.
\begin{table}[h]
\caption{Systematic frequency shifts and uncertainties of the Yb optical lattice clock.}  
	\label{clockbudget}
	\begin{center} 
\begin{tabular}{lrr}
\hline
Effect & Shift ($\times10^{-17}$) & Uncertainty ($\times10^{-17}$)\\
\hline
Lattice light shift & $14.9$ & 34.0 \\
Blackbody radiation & $-245.8$ & $10.4$ \\
Density &$-8.5$ & $5.3$ \\
Second order Zeeman & $-5.4$ & $0.3$ \\
Probe light shift & $1.1$ &$4.2$\\
Servo error & $1.3$ &$3.2$\\
AOM switching & $-$& $1$\\
\hline
Total  & $-242.4$& $36.3$ \\
\hline
\end{tabular}
\end{center}
\end{table}

\subsection{Lattice light shift}
\label{latticelightshiftsection}
\begin{figure}[h]
\begin{center}
\includegraphics[width=9cm,bb=0 50 910 716]{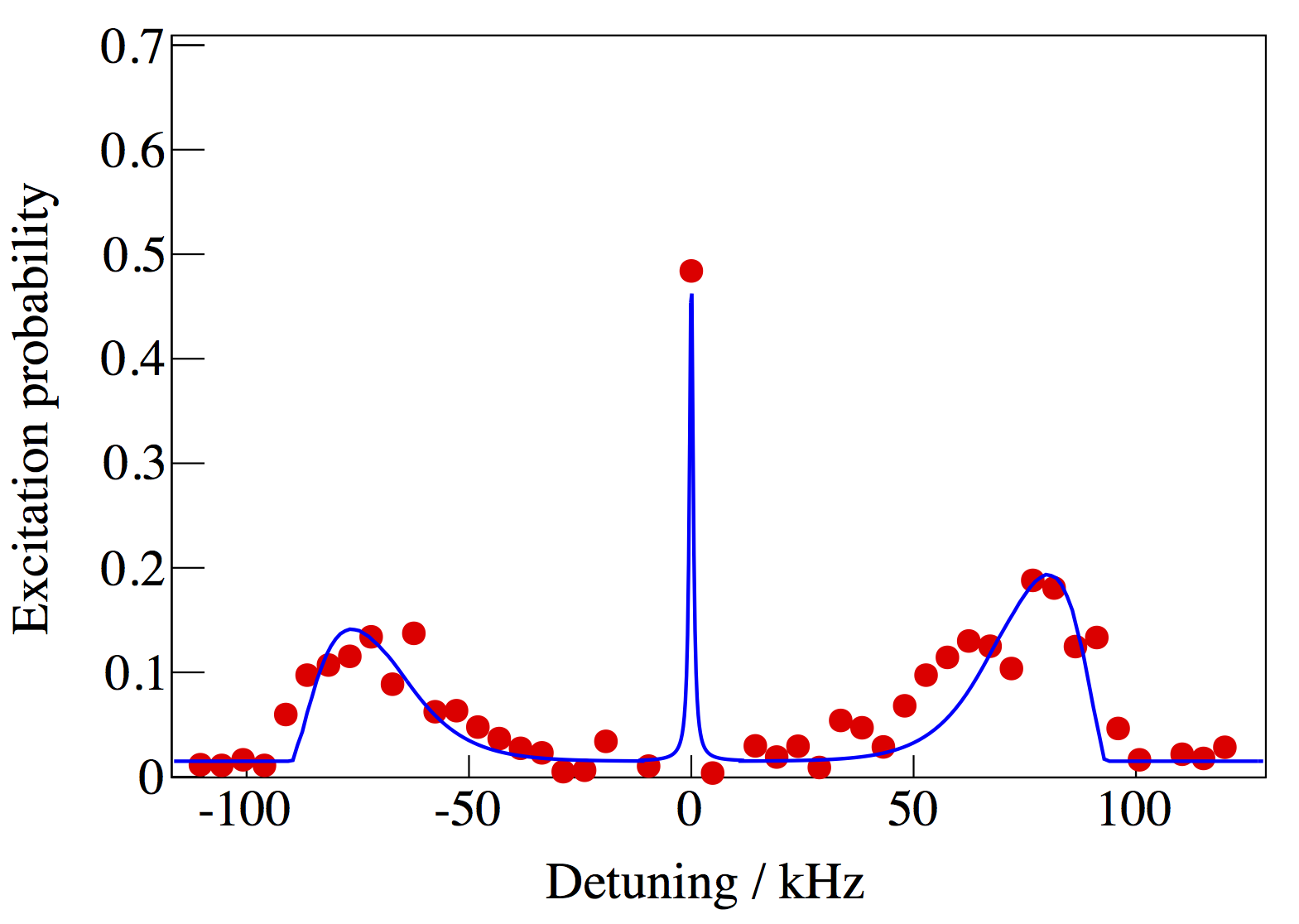}
\end{center}
\caption{Sideband spectrum of the clock transition. The solid blue line shows a fit of a combined function for the carrier and sidebands \cite{Blatt2009}.}
\label{sidebandspectrum}
\end{figure}

We observed a sideband spectrum to investigate the trap condition (see Fig. \ref{sidebandspectrum}). The spectrum was fitted with a combined profile of a Lorentzian function for the carrier and a function given in Ref. \cite{Blatt2009} for the sideband. From the fitting, the trap frequency in the longitudinal direction (parallel to the propagation of the lattice beam) was deduced to be 94(1) kHz corresponding to a trap potential depth of $U_{0}=537(12)E_{\mathrm{r}}$. The temperature of the atoms was estimated to be 12(4) \nolinebreak $\mathrm{\mu}$K from the ratio between the heights of the red and blue sidebands \cite{Blatt2009}. This ratio also provides the average vibrational quantum number of $\Braket{n}=2.2(7)$. 

\begin{figure}[h]
\begin{center}
\includegraphics[width=8cm,bb=0 50 910 716]{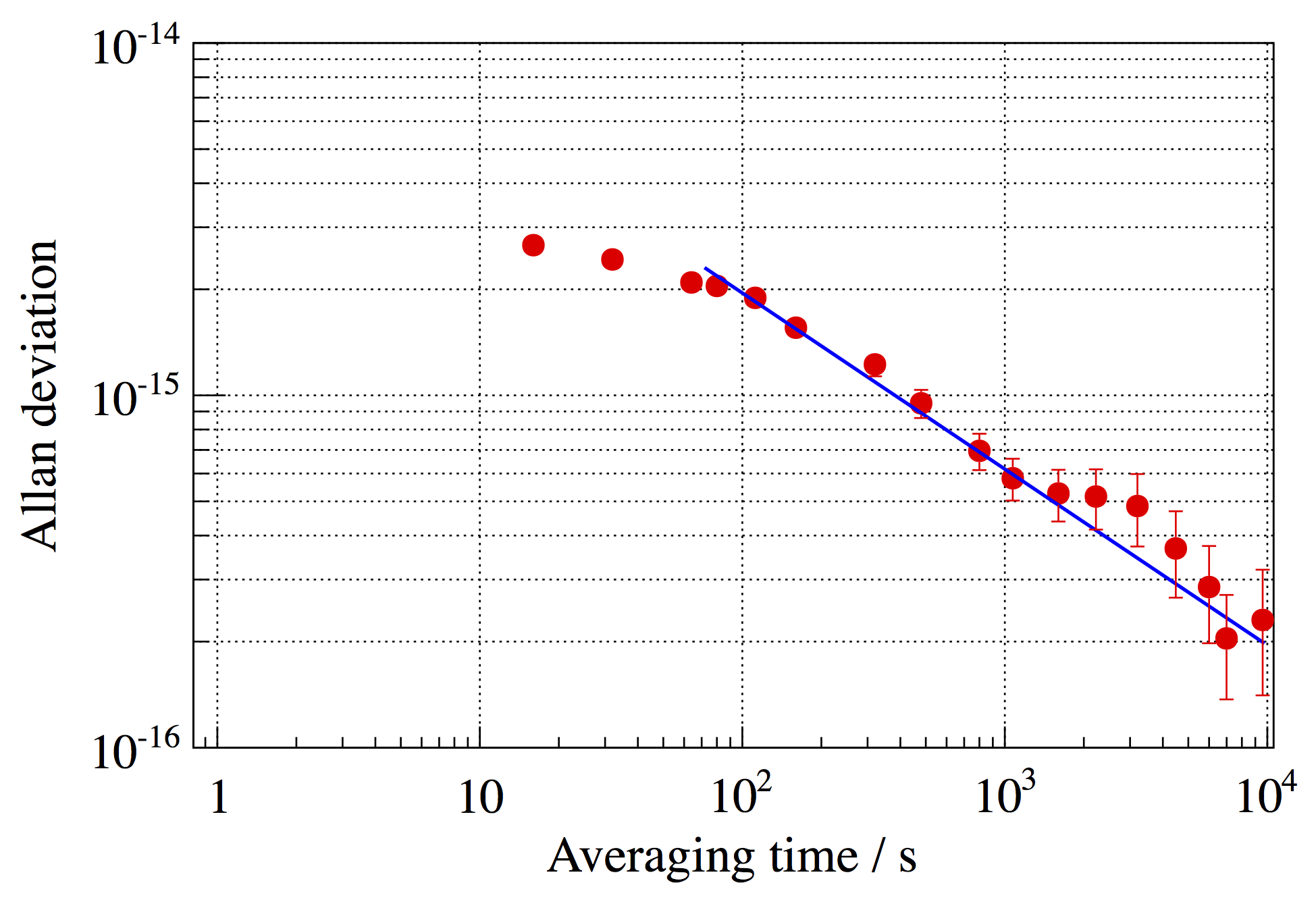}
\end{center}
\caption{Allan deviation of the interleaved measurement as a function of the averaging time $\tau$, which was calculated from the frequency shift data due to the difference between the trap potential depth of two values. The solid blue line indicates a slope of $2.0\times10^{-14}/\sqrt{\tau}$.}
\label{interleavedeviation}
\end{figure}

\begin{figure}[h]
\begin{center}
\includegraphics[width=9cm,bb=0 50 910 716]{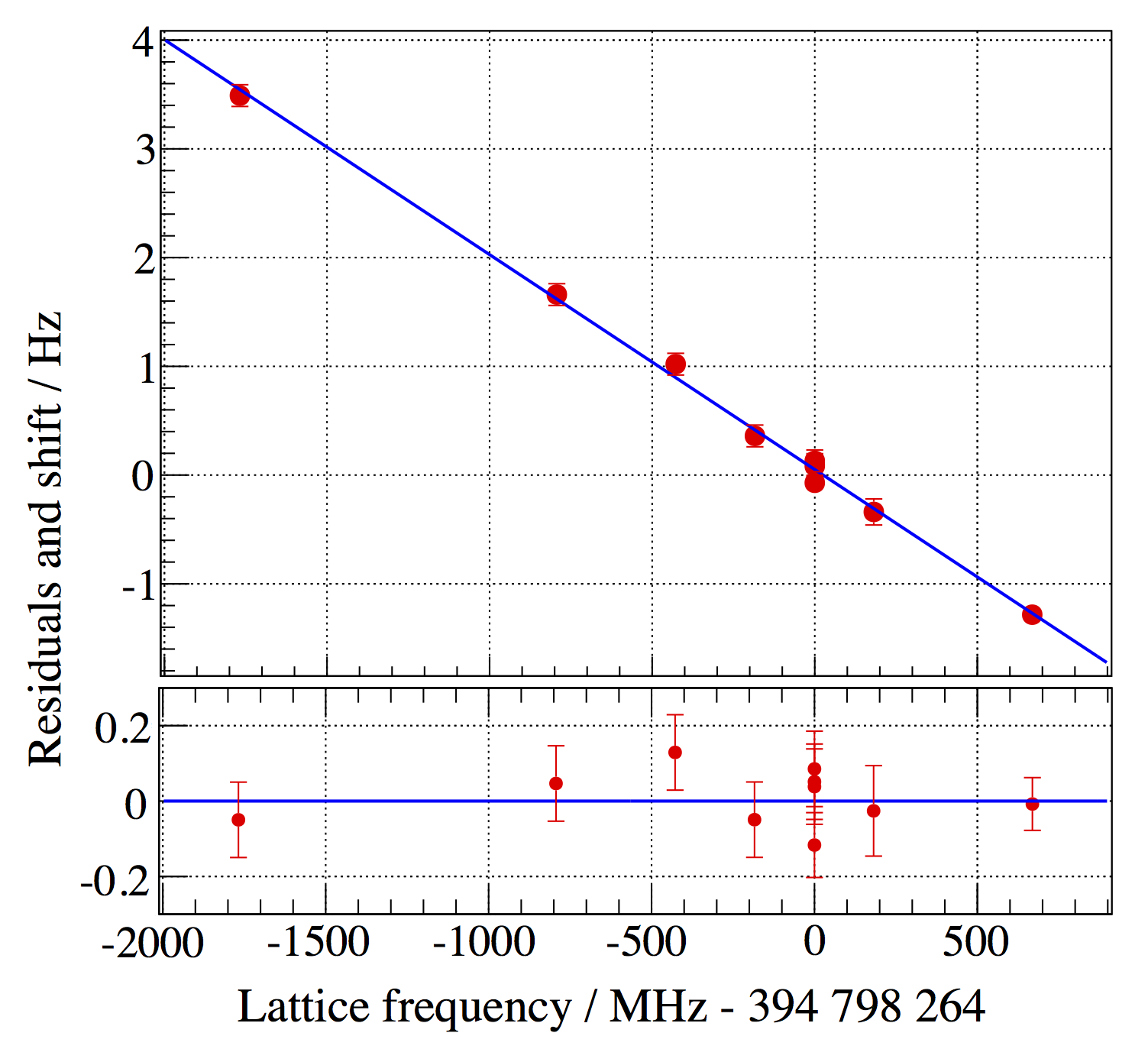}
\end{center}
\caption{Measured light shift as a function of the frequency of the lattice laser obtained by varying the trap potential depth between $U^{\mathrm{L}}_{0}=537(12)E_\mathrm{r}$ and $U^{\mathrm{H}}_{0}=670(13)E_\mathrm{r}$. The solid blue line indicates the fit of a light shift model  \cite{Katori2015,Nemitz2016} (see text). The frequency of $394\,798\,264$ MHz is our operating lattice frequency.}
\label{latticelightshift}
\end{figure}

To evaluate the lattice-induced light shift, we employed an interleaved scheme in which the frequency shift is measured by the alternating stabilization of the clock laser to the atomic transition with two different experimental parameters. Figure \ref{interleavedeviation} shows the Allan deviation of the interleaved measurement as a function of the averaging time $\tau$, which is calculated from the light shift data due to the difference in the trap depth between $U^{\mathrm{L}}_{0}=537(12)E_\mathrm{r}$ and $U^{\mathrm{H}}_{0}=670(13)E_\mathrm{r}$. The stability of the interleaved measurement was improved with a slope of $2.0\times10^{-14}/\sqrt{\tau}$. Figure \ref{latticelightshift} shows the measured light shift as a function of the frequency of the lattice laser. 

The lattice light shift is mainly attributed to the electric-dipole ($E1$) polarizability, which induces a linear shift against the detuning of the lattice laser frequency $\nu_{\mathrm{l}}$ from the $E1$ magic frequency $\nu_{E1}$ \cite{Katori2003}. The higher-order contribution to the light shift arises from the multipolar ($M1$ and $E2$) polarizability and hyperpolarizability that causes a nonlinear shift as a function of the trap depth \cite{Katori2015}. 
To analyze the measured shift data in Fig. \ref{latticelightshift}, we employed a model by RIKEN \cite{Katori2015,Nemitz2016} in which the light shift $\Delta \nu_{\mathrm{LS}}$ is given by 
\begin{eqnarray}
\Delta \nu_{\mathrm{LS}}&=&(a\Delta\nu - b)\Big(\Braket{n}+\frac{1}{2}\Big)\Big(U_{\mathrm{e}}/E_\mathrm{r}\Big)^{1/2}\nonumber\\
&&+d(2\Braket{n}+1)\Big(U_{\mathrm{e}}/E_{\mathrm{r}}\Big)^{3/2}\nonumber\\
&&-\Big\{a\Delta \nu + \frac{3}{4}d\Big(2\Braket{n}^{2}+2\Braket{n}+1\Big)\Big\}U_{\mathrm{e}}/E_\mathrm{r}\nonumber\\
&&-d\Big(U_{\mathrm{e}}/E_{\mathrm{r}}\Big)^2,
\label{lightshifteq}
\end{eqnarray}
where $\Delta\nu=\nu_{\mathrm{l}}-\nu_{E1}$, and $U_{e}=\xi U_{0}$ is the effective trap depth taking into account the atomic distribution in the radial (perpendicular) direction of the lattice beam \cite{Yamanaka2015}. The coefficient $a$ corresponds to the slope of the $E1$ polarizability per $\Delta\nu$, while $b$ and $d$, respectively, denote the multipolar polarizability and hyperpolarizability coefficients. In the present analysis, we used $b=-0.68(71)$ mHz taken from Ref. \cite{Nemitz2016}, and estimated $\xi=0.77(6)$, $\Braket{n}=2.2(7)$ at $U^{\mathrm{L}}_{0}$, and $\xi=0.78(6)$, $\Braket{n}=2.4(8)$ at $U^{\mathrm{H}}_{0}$ from the sideband spectra. 

\begin{figure}[h]
\begin{center}
\includegraphics[width=9cm,bb=0 50 910 716]{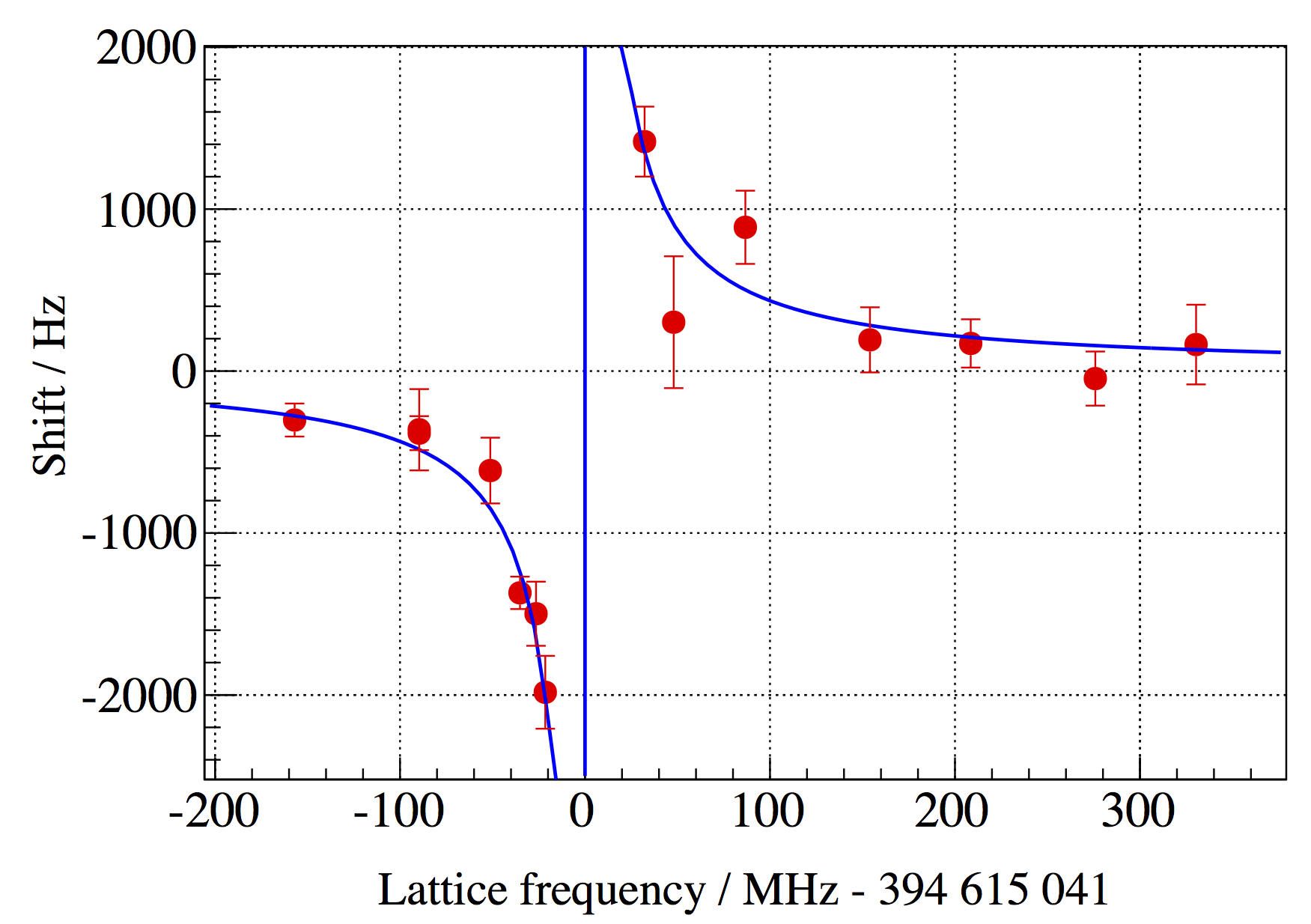}
\end{center}
\caption{Measured light shift as a function of the frequency of the lattice laser tuned close to the $\mathrm{6s6p^{3}P_{0}-6s8p^{3}P_{0}}$ two-photon resonance at $U^{\mathrm{H}}_{0}=670(13)E_\mathrm{r}$. The solid blue line indicates a fit of a dispersion line shape function.}
\label{hyperplot}
\end{figure}

To estimate the coefficient $d$, we evaluated the shift resulting from the hyperpolarizability at $U^{\mathrm{H}}_{0}$. This was carried out by measuring the light shifts at the lattice frequencies near the two-photon transitions $\mathrm{6s6p^{3}P_{0}-6s8p^{3}P_{0}}$, $\mathrm{6s8p^{3}P_{2}}$, and $\mathrm{6s5f^{3}F_{2}}$ at 759.71, 754.23, and 764.95 nm, respectively \cite{Barber2008} (see Fig. \ref{energydiagram}). The resonance frequencies of the two-photon transitions were confirmed by a significant reduction in the number of atoms in the $\mathrm{6s6p^{3}P_{0}}$ state remaining after the clock excitation. Figure \ref{hyperplot} shows the measured shift as a function of the lattice laser frequency tuned close to the $\mathrm{6s6p^{3}P_{0}-6s8p^{3}P_{0}}$ two-photon resonance. This transition is closest to the magic wavelength at 759 nm and thus makes a dominant contribution to the hyperpolarizability shift. The data points in Fig. \ref{hyperplot} were fitted with a dispersion line shape function. In this analysis, a constant offset was subtracted from the data. By extrapolating the fit function to the magic wavelength, we obtained a light shift of 235(18) mHz. The contribution of the other resonances $\mathrm{6s6p^{3}P_{0}-6s8p^{3}P_{2}}$, $\mathrm{6s5f^{3}F_{2}}$ to the shift is much smaller, since these resonances are far from the magic wavelength. Each of the resonances was observed with two hyperfine components \cite{Barber2008}. The data for each hyperfine component were similarly fitted with a dispersion line shape function and a light shift value was obtained by extrapolating the fit function to the magic wavelength. We did not observe statistically significant shifts for these resonances due to large linewidths and asymmetries of the observed clock spectra, resulting from frequency differences in the clock transition between different vibrational states \cite{Takamoto2003}. Therefore, we conservatively estimated a 1-standard-deviation lower shift limit of $-30$ mHz resulting from the $\mathrm{6s6p^{3}P_{0}-6s8p^{3}P_{2}}$ resonance, and a 1-standard-deviation upper shift limit of 16 mHz from the $\mathrm{6s6p^{3}P_{0}-6s5f^{3}F_{2}}$ resonance. In total, the hyperpolarizability shift was determined as $\Delta\nu_{\mathrm{HP}}=220(40)$ \nolinebreak mHz at $U^{\mathrm{H}}_{0}$. The coefficient $d$ was then calculated to be $-1.1(4)$ $\mathrm{\mu}$Hz to reproduce $\Delta\nu_{\mathrm{HP}}$ using Eq. (\ref{lightshifteq}). The uncertainty of $d$ mainly arose from the uncertainties of $\xi$ and $\Delta\nu_{\mathrm{HP}}$. The obtained $d$ coefficient agreed with the coefficient reported by RIKEN \cite{Nemitz2016} and its uncertainty was improved by a factor of 2 compared with that of RIKEN \cite{Nemitz2016}.

The data points in Fig. \ref{latticelightshift} were fitted with the difference $\Delta\nu_{\mathrm{LS}}(U^{\mathrm{H}}_{0})-\Delta\nu_{\mathrm{LS}}(U^{\mathrm{L}}_{0})$, treating $a$ and $\nu_{E1}$ as free parameters to be deduced from the fit. We employed a Monte-Carlo approach to take account of the uncertainties of the fixed parameters $b$, $U_{\mathrm{e}}$, $\Braket{n}$, and $d$. We numerically generated these parameters according to Gaussian distributions and repeated the fittings with different permutations of the parameters. The root mean square deviation of the deduced parameter ($a$ or $\nu_{E1}$) was calculated and quadratically added to the statistical uncertainty. We obtained $a=0.017(7)$ mHz MHz$^{-1}$ and $\nu_{E1}=394\,798\,247(26)$ MHz. The obtained $a$ coefficient was in good agreement with the coefficient reported by RIKEN \cite{Nemitz2016}. The $\nu_{E1}$ value agreed with the $E1$ magic frequencies of recent measurements \cite{Kim2017,Nemitz2016,Pizzocaro2017,Brown2017}. The light shift $\Delta\nu_{\mathrm{LS}}$ was calculated to be $77(176)$ mHz with a fractional uncertainty of $3.4\times10^{-16}$ under our operating conditions of $\nu_{\mathrm{l}}=394\,798\,264$ MHz and $U_{0}=537(12)E_\mathrm{r}$. The uncertainty of $\Delta\nu_{\mathrm{LS}}$ was mostly due to the relatively high trap depth resulting from the temperature of the atoms in the second-stage MOT and the uncertainty of $\xi$, and was not limited by the statistical uncertainty of the interleaved measurement.

\subsection{Blackbody radiation shift}
To estimate the blackbody radiation shift, we measured the temperature of the vacuum chamber and bobbins for the MOT coils in a vacuum, which are directly seen by the atoms. The bobbins were made of copper and supported by a copper rod. The rod penetrated the vacuum chamber into the atmosphere. The heat generated in the MOT coils was conducted through this rod to the atmosphere. To adjust the temperature of the bobbins close to room temperature, the atmosphere edge of the rod was cooled using a Peltier device. The temperature was measured using PT100 resistors on a) the cooled edge of the rod, b) one point on the bobbin furthest from the atmosphere, and c) four points on the stainless steel vacuum chamber including the point closest to and the point furthest from the Zeeman slower coil. We found that the temperatures of the measured points reached stable values $\sim1.5$ hours after starting the clock cycle, due to the heat arising from the MOT and Zeeman slower coils. 

After the thermal equilibrium had been reached, the maximum and minimum temperatures were found to be $T_{\mathrm{max}}=305$ K at the MOT bobbin inside the vacuum chamber and $T_{\mathrm{min}}=295$ K at the chamber furthest from the Zeeman slower coil, respectively. This indicates that the effective temperature seen by the atoms lies between two bounds $T_{\mathrm{min}}$ and $T_{\mathrm{max}}$. Since we did not have further knowledge about the spatial distribution of the temperature, we assumed a rectangular probability distribution of the effective temperature between $T_{\mathrm{min}}$ and $T_{\mathrm{max}}$, according to Refs. \cite{BIPM2008,Pizzocaro2017,Lodewyck2016}. With this assumption, the effective temperature was calculated to be $(T_{\mathrm{max}}+T_{\mathrm{min}})/2=300$ K with an uncertainty of $(T_{\mathrm{max}}-T_{\mathrm{min}})/\sqrt{12}=3$ K. The blackbody radiation shift was then estimated to be $-1.273(54)$ Hz with a fractional uncertainty of $1.0\times10^{-16}$ using the coefficient reported by NIST \cite{Sherman2012,Beloy2014} and the fact that the shift scales as $T^{4}$.

The contribution of the radiation from the atomic oven was also considered. Blackbody radiation photons emerging from the oven heated at 653 K are emitted through an aperture with a diameter of 6 mm and at a distance of 300 mm from the atoms. The photons reach the atoms directly or after being multiply scattered on the wall of the vacuum system. We employed a simple and conservative model, similarly to Refs. \cite{Middelmann2010,Lodewyck2016}, in which the atoms are located at the center of a stainless steel sphere with a radius of 300 mm and an emissivity of 0.1. The blackbody radiation photons at 653 K are provided from a small portion of the sphere's surface with its area equal to the aperture of the oven. The relative frequency shift due to the directly reached or multiply scattered photons was calculated to be $-1\times10^{-17}$ using a method of Ref. \cite{Middelmann2010}. This shift value was taken as an uncertainty and quadratically added to the above uncertainty estimated by the temperature measurement.

\begin{figure}[h]
\begin{center}
\includegraphics[width=9cm,bb=0 50 910 716]{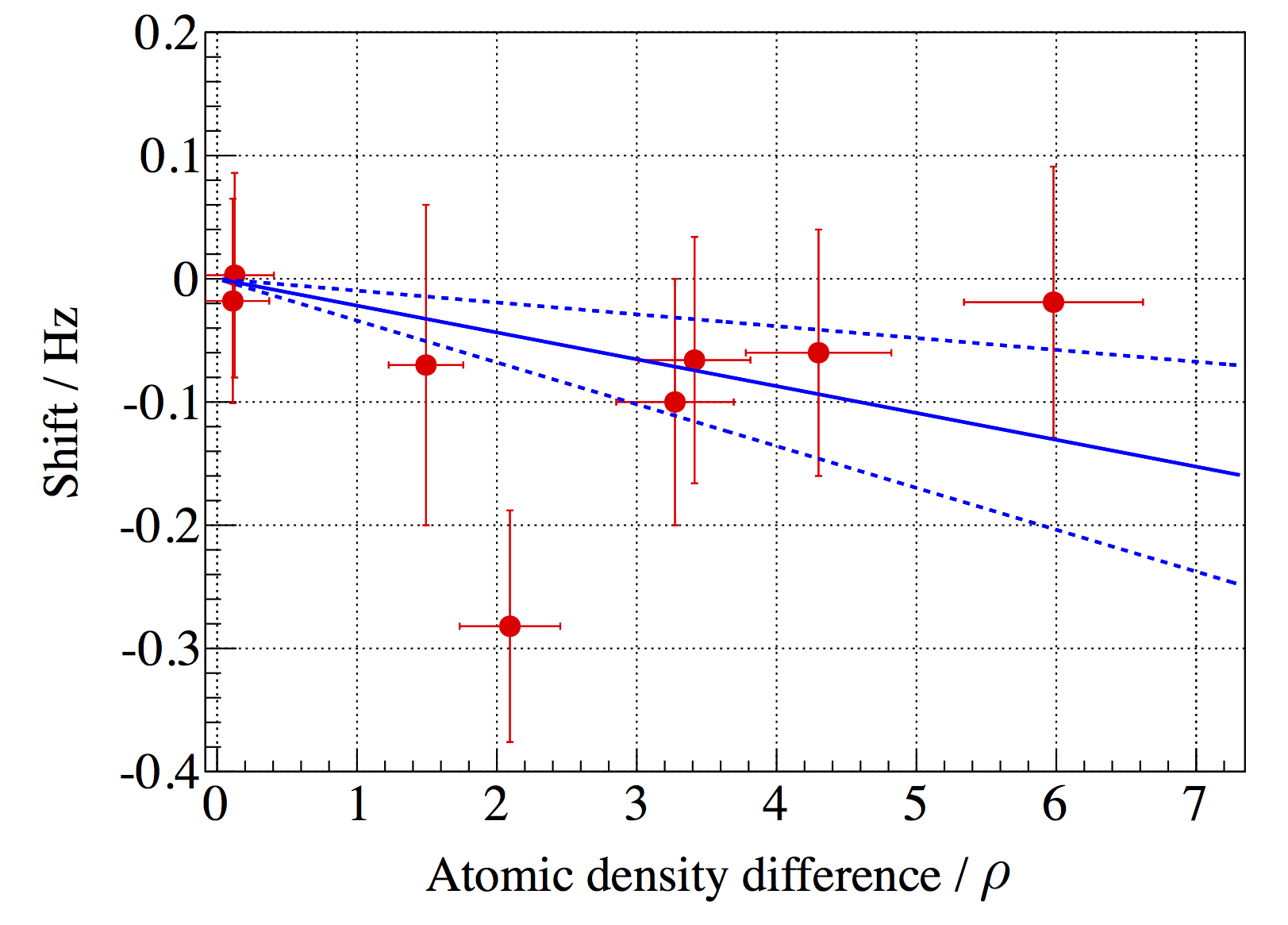}
\end{center}
\caption{Measured density shift as a function of the atomic density difference ($\rho\sim10^{14}$ m$^{-3}$). The solid and dashed blue lines indicate a linear fit and its uncertainty, respectively.}
\label{densityshift}
\end{figure}

\subsection{Density shift}
The density shift caused by cold atomic collisions was evaluated with an interleaved measurement, changing the number of atoms trapped in the optical lattice. Figure \ref{densityshift} shows the measured shift as a function of the atomic density difference. The number of atoms was estimated from the fluorescence signal with (a) the quantum efficiency of the EMCCD, (b) the photon scattering rate calculated with the intensity and detuning of the 399-nm laser, (c) the solid angle of the detection system, and (d) the decay time of the florescence signal. The trap volume of the lattice was calculated to be the order of $10^{-12}$ m$^{3}$ from a radius of $\sim20$ $\mathrm{\mu}$m and a longitudinal length of $\sim2$ mm that was estimated from the Rayleigh length and the size of the 566-nm MOT. By fitting the measured data with a linear function, we obtained a slope of $-22(12)$ mHz/$\rho$ where $\rho\sim10^{14}$ m$^{-3}$ was introduced as an arbitrary reference density to factor out the uncertainty of the estimated density. For our operating condition of $2.0(6)\rho$, the density shift was $-44(27)$ mHz with a relative uncertainty of $5.3\times10^{-17}$.

\subsection{Zeeman shift}
\begin{figure}[h]
\begin{center}
\includegraphics[width=9cm,bb=0 50 910 716]{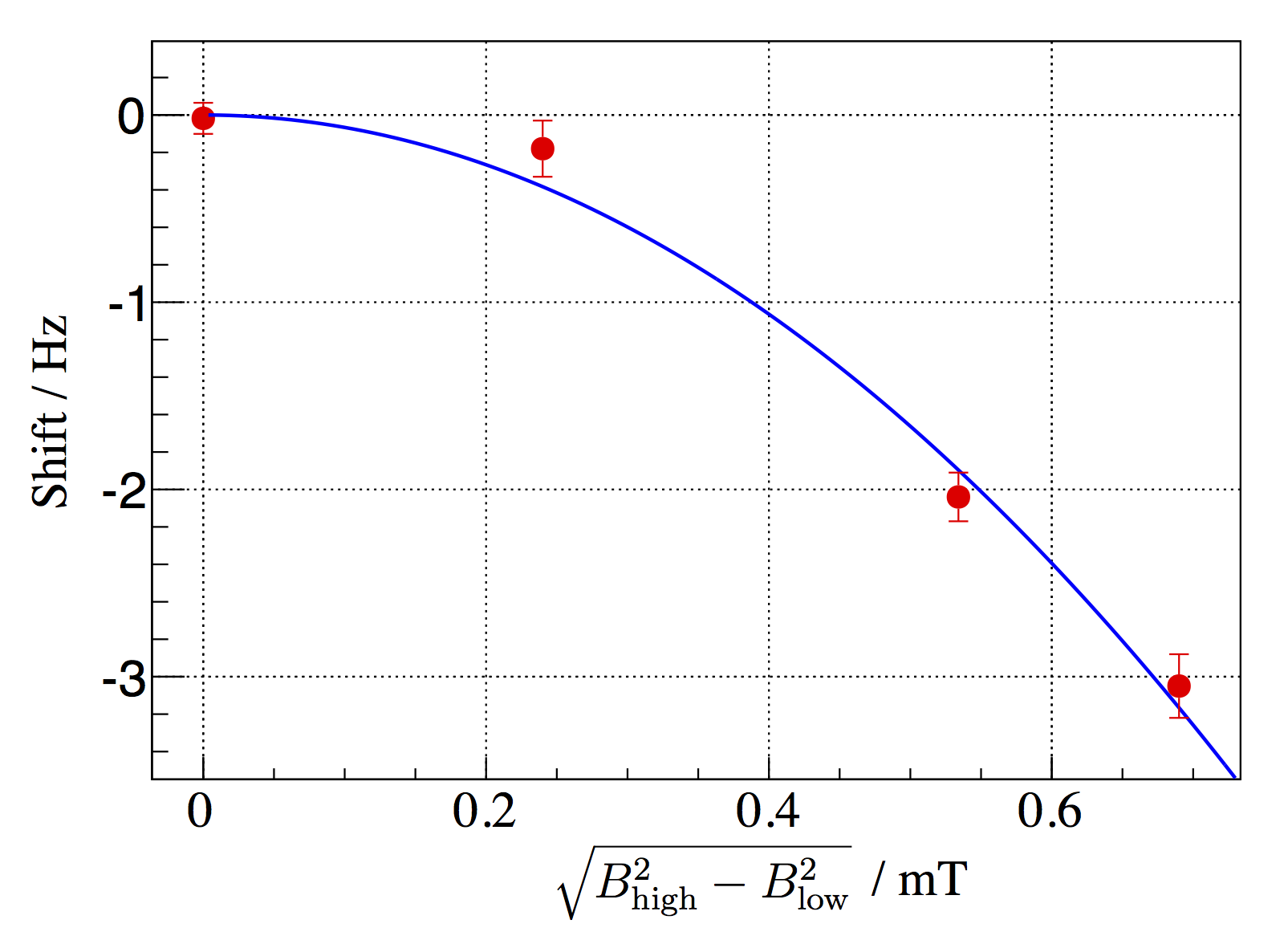}
\end{center}
\caption{Measured second-order Zeeman shift as a function of $\sqrt{B_{\mathrm{high}}^{2}-B_{\mathrm{low}}^{2}}$, where $B_{\mathrm{high}}$ and $B_{\mathrm{low}}$ are high and low external bias magnetic fields. The solid blue line indicates the fit of a quadratic function.}
\label{secondzeemanshift}
\end{figure}

The first-order Zeeman effect, which gives a linear shift against the magnetic field, was removed by locking the clock laser to the mean frequency of the two $\pi$ transitions as described in Sect. \ref{clockoperationsection}. The quadratic second-order Zeeman shift was evaluated with an interleaved measurement in which the external bias magnetic field was varied. Figure \ref{secondzeemanshift} shows the measured second-order Zeeman shift. The shift $\Delta \nu_{\mathrm{Z}}$ obtained in the interleaved scheme between high and low magnetic fields $B_{\mathrm{high}}$ and $B_{\mathrm{low}}$ can be written as
$\Delta\nu_{\mathrm{Z}}=a_{\mathrm{Z}}(B_{\mathrm{high}}^{2}-B_{\mathrm{low}}^{2})$,
where $a_{\mathrm{Z}}$ denotes the coefficient of the second-order Zeeman shift. A quadratic dependence of $\Delta \nu_{\mathrm{Z}}$ should therefore be observed as a function of $\sqrt{B_{\mathrm{high}}^{2}-B_{\mathrm{low}}^{2}}$. The magnetic fields were calculated from the Zeeman splittings and the coefficient of the first-order Zeeman shift \cite{Lemke2009}. The Zeeman splitting and its uncertainty was determined from the mean and standard deviation of the splitting data obtained in the interleaved measurement. The uncertainty of the Zeeman splitting was typically 4 \nolinebreak Hz corresponding to the magnetic field uncertainty of 1 \nolinebreak $\mathrm{\mu}$T. The contribution of the vector light shift \cite{Lemke2009} to the observed splitting was considered to be negligibly small compared with the splitting uncertainty of 4 \nolinebreak Hz in our experimental condition ($E_{\mathrm{L}}\parallel$ $B_{\mathrm{Bias}}$, see Fig. \ref{setupfigure}). We obtained $a_{\mathrm{Z}}=-6.6(3)$ \nolinebreak Hz/mT$^{2}$ by applying a quadratic fit of the data in Fig. \ref{secondzeemanshift}. This coefficient was in good agreement with that reported by NIST \cite{Lemke2009} and ECNU \cite{Gao2018}, and its uncertainty was improved by a factor of about 3 compared with that of NIST\cite{Lemke2009}. The obtained coefficient can also be written as $-1.49(7)$ \nolinebreak $\mathrm{\mu}$Hz/Hz$^{2}$ when the second-order Zeeman shift is given as a function of the first-order Zeeman shift. For our operating condition of 65(1) $\mathrm{\mu}$T (i.e., the first-order Zeeman shift of 137(2) Hz), the second-order Zeeman shift was $-27.9(1.5)$ mHz with a relative uncertainty of $3\times10^{-18}$.

\subsection{Probe light shift}
The probe light shift resulting from the irradiation of the clock laser was calculated from the shift value of NIST \cite{Lemke2009}. During the present measurements, the duration $t_{\mathrm{\pi}}$ of the $\pi$ pulse was kept at 55 ms, whereas $t_{\mathrm{\pi}}=80$ ms in NIST \cite{Lemke2009}. The relative shift was estimated to be $1.1(4.2)\times10^{-17}$ using the fact that the laser intensity adjusted for the $\pi$ pulse scales as $t_{\mathrm{\pi}}^{-2}$.

\subsection{Servo error}
To estimate the servo error, we averaged the differences in the excitation probabilities between the high- and low-frequency spectral shoulders over the data obtained for $\sim2\times10^{5}$ s. This gave a relative shift of $1.3(3.2)\times10^{-17}$. 

\subsection{AOM switching}
The AOM for controlling the clock frequency was operated with a low RF power of $\sim15$ mW to reduce thermal effects. A resonant clock pulse was realized by detuning the frequency of the AOM by $\sim$100 kHz outside the probe time, while maintaining a constant RF power. We made no attempt to measure the shift caused by the AOM switching. We therefore set a conservative uncertainty of $1\times10^{-17}$ resulting from the AOM switching based on previous measurements obtained under similar experimental conditions \cite{Nemitz2016,Pizzocaro2017}.

\subsection{DC Stark shift}
A dc Stark shift can arise from a static electric field. The stainless steel vacuum chamber is connected to a ground potential and serves as a Faraday cage against external electric fields. The MOT coil bobbins inside a vacuum are connected to the vacuum chamber and thus to ground potential. A fused silica mirror with a quarter-wave plate for retro-reflecting the MOT beam is placed inside the vacuum chamber and 60 mm from the atoms. Several fused silica viewports are located at distances of 80 mm from the atoms. The contribution of possible electric charges trapped in the mirror or viewports to the frequency shift \cite{Lodewyck2012} was considered to be negligibly small, since their distances are large enough and the coil bobbins partially shield the atoms from residual electric fields. Therefore, we did not include the dc Stark effect in the uncertainty budget. 

\section{Discussions and Conclusions}
\label{discussionsection}
The systematic uncertainty of the Yb lattice clock was mainly limited by the lattice light shift and the blackbody radiation shift. The uncertainty of the lattice light shift was mostly due to the high trap depth $U_{0}=540E_\mathrm{r}$ and the uncertainty of the effective trap depth $\xi U_{0}$. The uncertainty of the light shift can be improved to the $10^{-17}$ level by reducing the trap depth to $<200E_\mathrm{r} \sim20$ $\mathrm{\mu}$K. This was difficult in the present experiment because of the high temperature of the atoms in the second-stage MOT ($\sim40$ \nolinebreak $\mathrm{\mu}$K). In the future, the temperature of the atoms will be reduced to near the Doppler limit ($\sim4$ $\mathrm{\mu}$K). The uncertainty of the blackbody radiation shift was limited by the large temperature uncertainty of 3 K. This arose from the heat generated in the MOT coils inside the vacuum chamber that were operated with the relatively long duration of 1.5 s in the first-stage MOT against the clock cycle time of 2 s. The ratio of the MOT duration to the clock cycle time will be reduced by increasing the power of the MOT beam at 399 nm and improving the transfer efficiency of atoms from the MOT to the lattice. This will improve the uncertainty of the blackbody radiation shift to the $10^{-17}$ level. 

The previous determination of the hyperporlizability shift coefficient $d$ at RIKEN \cite{Nemitz2016} was based on a measurement of the hyperpolarizability shift performed at NIST \cite{Barber2008} and limited by a large uncertainty in the estimation of the effective trap depth $\xi U_{0}$ used in the NIST measurement. The improvement in the uncertainty of $d$ in the present work was due to the measurement of both the hyperpolarizability shift and $\xi U_{0}$ using the same experimental setup. The light shift model by RIKEN has been employed at RIKEN \cite{Nemitz2016}, KRISS\cite{Kim2017}, and INRiM \cite{Pizzocaro2017}. Our coefficient $d$ should reduce the total systematic uncertainties of Yb lattice clocks at RIKEN and INRiM when calculated with the reported trap depths $\xi U_{0}$, since their uncertainties are mainly limited by the light shift uncertainties. 

The lattice-induced light shift has recently been evaluated at the $10^{-18}$ level at NIST \cite{Brown2017}. The coefficients for the light shift including the hyperpolaizability shift are provided with a simple model that does not include a parameter representing the atomic temperature in the lattice. This model is based on the fact that the atomic temperature is proportional to the trap depth, which has been observed at NIST. Since we did not observe a significant difference of the temperature by varying the trap depth between $U_{0}=290E_\mathrm{r}$ and $670E_\mathrm{r}$, we did not employ this model in the present evaluation.

The previous coefficient $a_{\mathrm{Z}}$ of the second-order Zeeman shift determined at NIST \cite{Lemke2009} has been used in the calculation of this shift at NIST \cite{Lemke2009}, RIKEN \cite{Nemitz2016}, INRiM \cite{Pizzocaro2017}, and KRISS \cite{Kim2017}. The contribution of this shift to the total systematic uncertainty is the largest at KRISS and the second largest at RIKEN. The present coefficient $a_{\mathrm{Z}}$ should improve the total uncertainties of Yb clocks at those institutes when calculated with the reported magnetic fields.

In conclusion, we have evaluated the systematic frequency shifts of an Yb lattice clock with a total uncertainty of $3.6\times10^{-16}$, which was mostly limited by the lattice-induced light shift and the blackbody radiation shift. 
In this evaluation, the uncertainties of the shift coefficients for the hyperpolarizability shift and the second-order Zeeman shift were improved compared with the values obtained in previous experiments \cite{Nemitz2016,Lemke2009,Barber2008}. In the future, we will compare the Yb clock with our Sr lattice clock \cite{Akamatsu2014ape,Tanabe2015} to measure the frequency ratio Yb/Sr.

\section*{Acknowledgment}
We are indebted to H. Katori, M. Takamoto, and H. Imai for providing information on their vacuum systems and valuable comments. We thank S. Okubo for the development of the fiber combs and for technical assistance. We are grateful to T. Kohno for the loan of the 556-nm laser system. We wish to dedicate this paper to the memory of A.
Onae. 
\ifCLASSOPTIONcaptionsoff
  \newpage
\fi

\end{document}